\documentstyle[twoside]{article}

\catcode`\@=11
\long\def\@makefntext#1{
\protect\noindent \hbox to 3.2pt {\hskip-.9pt
$^{{\eightrm\@thefnmark}}$\hfil}#1\hfill}               

\def\@makefnmark{\hbox to 0pt{$^{\@thefnmark}$\hss}}    

\def\ps@myheadings{\let\@mkboth\@gobbletwo
\def\@oddhead{\hbox{}
\rightmark\hfil\eightrm\thepage}
\def\@oddfoot{}\def\@evenhead{\eightrm\thepage\hfil
\leftmark\hbox{}}\def\@evenfoot{}
\def\sectionmark##1{}\def\subsectionmark##1{}}

\oddsidemargin=\evensidemargin
\newcounter{sectionc}\newcounter{subsectionc}\newcounter{subsubsectionc}
\renewcommand{\section}[1] {\vspace{12pt}\addtocounter{sectionc}{1}
\setcounter{subsectionc}{0}\setcounter{subsubsectionc}{0}\noindent
        {\tenbf\thesectionc. #1}\par\vspace{5pt}}
\renewcommand{\subsection}[1] {\vspace{12pt}\addtocounter{subsectionc}{1}
      \setcounter{subsubsectionc}{0}\noindent
      {\bf\thesectionc.\thesubsectionc.{\kern1pt \bfit #1}}\par\vspace{5pt}}
\renewcommand{\subsubsection}[1]
      {\vspace{12pt}\addtocounter{subsubsectionc}{1}
      \noindent{\tenrm\thesectionc.\thesubsectionc.\thesubsubsectionc.
      {\kern1pt \tenit #1}}\par\vspace{5pt}}
\newcommand{\nonumsection}[1] {\vspace{12pt}\noindent{\tenbf #1}
        \par\vspace{5pt}}

\newcounter{appendixc}
\newcounter{subappendixc}[appendixc]
\newcounter{subsubappendixc}[subappendixc]
\renewcommand{\thesubappendixc}{\Alph{appendixc}.\arabic{subappendixc}}
\renewcommand{\thesubsubappendixc}
        {\Alph{appendixc}.\arabic{subappendixc}.\arabic{subsubappendixc}}

\renewcommand{\appendix}[1] {\vspace{12pt}
        \refstepcounter{appendixc}
        \setcounter{figure}{0}
        \setcounter{table}{0}
        \setcounter{lemma}{0}
        \setcounter{theorem}{0}
        \setcounter{corollary}{0}
        \setcounter{definition}{0}
        \setcounter{equation}{0}
        \renewcommand{\thefigure}{\Alph{appendixc}.\arabic{figure}}
        \renewcommand{\thetable}{\Alph{appendixc}.\arabic{table}}
        \renewcommand{\theappendixc}{\Alph{appendixc}}
        \renewcommand{\thelemma}{\Alph{appendixc}.\arabic{lemma}}
        \renewcommand{\thetheorem}{\Alph{appendixc}.\arabic{theorem}}
        \renewcommand{\thedefinition}{\Alph{appendixc}.\arabic{definition}}
        \renewcommand{\thecorollary}{\Alph{appendixc}.\arabic{corollary}}
        \renewcommand{\theequation}{\Alph{appendixc}.\arabic{equation}}
        \noindent{\tenbf Appendix \theappendixc #1}\par\vspace{5pt}}
\newcommand{\subappendix}[1] {\vspace{12pt}
        \refstepcounter{subappendixc}
        \noindent{\bf Appendix \thesubappendixc. {\kern1pt \bfit #1}}
        \par\vspace{5pt}}
\newcommand{\subsubappendix}[1] {\vspace{12pt}
        \refstepcounter{subsubappendixc}
        \noindent{\rm Appendix \thesubsubappendixc. {\kern1pt \tenit #1}}
        \par\vspace{5pt}}

\newcommand{\smalllineskip}{\baselineskip=10pt}

\def\eightcirc{
\begin{picture}(0,0)
\put(4.4,1.8){\circle{6.5}}
\end{picture}}
\def\eightcopyright{\eightcirc\kern2.7pt\hbox{\eightrm c}}

\def\abstracts#1#2#3{{
        \centering{\begin{minipage}{4.5in}\baselineskip=10pt\footnotesize
        \parindent=0pt #1\par
        \parindent=15pt #2\par
        \parindent=15pt #3
        \end{minipage}}\par}}

\renewenvironment{thebibliography}[1]
        {\frenchspacing
         \ninerm\baselineskip=11pt
         \begin{list}{\arabic{enumi}.}
        {\usecounter{enumi}\setlength{\parsep}{0pt}
         \setlength{\leftmargin 12.7pt}{\rightmargin 0pt} 
         \setlength{\itemsep}{0pt} \settowidth
        {\labelwidth}{#1.}\sloppy}}{\end{list}}

\newcounter{itemlistc}
\newcounter{romanlistc}
\newcounter{alphlistc}
\newcounter{arabiclistc}

\newcommand{\fcaption}[1]{
        \refstepcounter{figure}
        \setbox\@tempboxa = \hbox{\footnotesize Fig.~\thefigure. #1}
        \ifdim \wd\@tempboxa > 5in
           {\begin{center}
        \parbox{5in}{\footnotesize\smalllineskip Fig.~\thefigure. #1}
            \end{center}}
        \else
             {\begin{center}
             {\footnotesize Fig.~\thefigure. #1}
              \end{center}}
        \fi}

\newcommand{\tcaption}[1]{
        \refstepcounter{table}
        \setbox\@tempboxa = \hbox{\footnotesize Table~\thetable. #1}
        \ifdim \wd\@tempboxa > 5in
           {\begin{center}
        \parbox{5in}{\footnotesize\smalllineskip Table~\thetable. #1}
            \end{center}}
        \else
             {\begin{center}
             {\footnotesize Table~\thetable. #1}
              \end{center}}
        \fi}

\def\@citex[#1]#2{\if@filesw\immediate\write\@auxout
        {\string\citation{#2}}\fi
\def\@citea{}\@cite{\@for\@citeb:=#2\do
        {\@citea\def\@citea{,}\@ifundefined
        {b@\@citeb}{{\bf ?}\@warning
        {Citation `\@citeb' on page \thepage \space undefined}}
        {\csname b@\@citeb\endcsname}}}{#1}}

\newif\if@cghi
\def\cite{\@cghitrue\@ifnextchar [{\@tempswatrue
        \@citex}{\@tempswafalse\@citex[]}}
\def\citelow{\@cghifalse\@ifnextchar [{\@tempswatrue
        \@citex}{\@tempswafalse\@citex[]}}
\def\@cite#1#2{{$\null^{#1}$\if@tempswa\typeout
        {IJCGA warning: optional citation argument
        ignored: `#2'} \fi}}

\def\@refcitex[#1]#2{\if@filesw\immediate\write\@auxout
        {\string\citation{#2}}\fi
\def\@citea{}\@refcite{\@for\@citeb:=#2\do
        {\@citea\def\@citea{, }\@ifundefined
        {b@\@citeb}{{\bf ?}\@warning
        {Citation `\@citeb' on page \thepage \space undefined}}
        \hbox{\csname b@\@citeb\endcsname}}}{#1}}

\def\@refcite#1#2{{#1\if@tempswa\typeout
        {IJCGA warning: optional citation argument
        ignored: `#2'} \fi}}

\def\refcite{\@ifnextchar[{\@tempswatrue
        \@refcitex}{\@tempswafalse\@refcitex[]}}

\def\pmb#1{\setbox0=\hbox{#1}
        \kern-.025em\copy0\kern-\wd0
        \kern.05em\copy0\kern-\wd0
        \kern-.025em\raise.0433em\box0}

\def\fnt#1#2{\footnotetext{\kern-.3em
        {$^{\mbox{\scriptsize #1}}$}{#2}}}

\font\tenrm=cmr10
\font\tenit=cmti10
\font\tenbf=cmbx10
\font\bfit=cmbxti10 at 10pt
\font\ninerm=cmr9

\font\eightrm=cmr8

\def\qed{\hbox{${\vcenter{\vbox{                      
   \hrule height 0.4pt\hbox{\vrule width 0.4pt height 6pt
   \kern5pt\vrule width 0.4pt}\hrule height 0.4pt}}}$}}

\begin{document}


\vspace*{0.88truein}

\centerline{\bf Comment on ``Dependence of Gravitational Action on
Chemical Composition:}
\vspace*{0.035truein}
\centerline{\bf  New Series of Experiments" by M. Nanni in Apeiron vol.7,
p.  195 (2000)} \vspace*{0.035truein}

\vspace*{0.37truein}
\centerline{\footnotesize Rumen I. Tzontchev and Andrew E. Chubykalo}

\centerline{\footnotesize \it
Escuela de F\'{\i}sica, Universidad Aut\'onoma de Zacatecas}
\baselineskip=10pt
\centerline{\footnotesize \it
Apartado Postal C-580\, Zacatecas 98068, ZAC., M\'exico}


\baselineskip 5mm

\vspace*{0.21truein}

\abstracts{In our comment we show that the application of appropriated
statistical methods to the results of the author proves that the author in
his article has not been able to reach his goal.}{}{}


\bigskip

$$$$

In article [1] the results of a very interesting fundamental experiment
are described. The objective of the experiment is to statistically
demonstrate that the folowing equation is reliable
$$
M=\frac{W_i}{W_k}({\rm Torino}) -  \frac{W_i}{W_k}({\rm Plateau\;
Ros{\acute a}})\neq 0
$$
where $W_i$ and $W_k$ are correspondingly the weights of samples of two
materials of different chemical compositions, measured in the city of
Torino (180m above sea level) and in Plateau Ros\'a (3480m above sea
level). This would seriously question the validity of the Weak
Equivalence Principle (WEP). Our purpose is to show that the author has
not been able to reach his goal in his article. With that purpose a
standart statistical processing of the author's presented results has
been completed in [1], using the same symbols. The following relationships
have been used [2-4] (the letter ``$A$" corresponds  to Torino, the letter
``$B$" corresponds to Plateau Ros\'a):
$$
\Delta W= \sqrt{[SDA\cdot t(P,n)]^2+\Delta^2_d};
$$
\bigskip
$$
\Delta\left[\frac{W_i}{W_k}\right]=
\frac{\overline{W}_k\cdot \Delta W_i+\overline{W}_i\cdot\Delta
W_k}{\overline{W}_k^2};
$$
\bigskip
$$
\Delta M=\Delta\left[\frac{W_i}{W_k}(A)\right]+
\Delta\left[\frac{W_i}{W_k}(B)\right],
$$
where $\Delta W$ is the experimental error of a series of measurements  of
the weight of a sample under certain conditions; $SDA$ is Standard
Deviation Average for the same weight; $t(P;n)$ it is the Student's
coefficient to confidence probability $P$ and number measurements $n$;
$\Delta_d$ is the scale error ($\Delta_d= 3\times 10^{-6}{\sl g})$;
$\overline{W}$ is the average of the corresponding sample weight; $\Delta
M$ determines the limits of the confidence interval $(M -\Delta M$, $M
+\Delta M)$.  With probability $P$ the exact value of magnitude $M$ is
located within this interval.

>From the results in [1] the accuracy  with which the experiment should be
carried out is seen, it is comparable with the accuracy of a metrological
experiment. For following, in the statistical processing of the 
experimental data, the requirements of a metrological experiment should be 
respected.  For that reason, a level of the confidence probability $P$ has 
been accepted as 0.999. On the other hand, the noted confidence 
probability is required for each experiment that aspires to demonstrate 
invalidity in a fundamental physical principle.  If the weight of a sample 
is measured 10 times in an experimental series and $P = 0.999$, the 
Student's coefficient is valued as $t(0.999,10) = 4.78$.  There are two 
possibilities:

1. If the digit ``0" is outside the confidence interval, it
can be confirmed with a probability of 0.999, that the exact value of
magnitude M is different from ``0".

2. If the digit "0" is inside the  confidence interval, nothing can be
deduced.

The deviation limit $\Delta M$, that is only due  to the scale error
$\Delta_d$, equals $4\times 10^{-6}$. Because of this, there is no reason
to consider those combinations of chemical substances, where
$M \leq 4\times 10^{-6}$, and only the cases where $M>4\times 10^{-6}$
will be dealt with. In Table 1 magnitude $M$ (calculated by M. Nanni),
$\Delta M$, the reliable interval and the relative error for several
chemical substance combinations have been presented:
$$$$

\begin{center}
\begin{tabular}{|c|c|c|c|c|c|}    \hline
{\bf Lead} &  Aluminium & Gold & Bronze & Silver & Brass-Sand \\  \hline
$M(\times 10^{-6})$ & 8 & 8 & 6 & 6 & 6 \\ \hline
$\Delta M(\times 10^{-6})$ & 9.07 & 8.11 & 9.42 & 8.79 & 6.26 \\ \hline
Confidence & (-1.07; & (-0.11; & (-3.42; & (-2.79; & (-0.26; \\
interval & 17.07) & 16.11) & 15.42) & 14.79) & 12.26) \\ \hline
$\frac{\Delta M}{M}\cdot 100\%$ & 113\% & 101\% & 157\% & 146\% & 104\% \\
\hline
\end{tabular}
\end{center}

\medskip

\begin{center}
Table 1
\end{center}

$$$$
It can be seen that digit ``0" participates  in all of the confidence
intervals. This clearly indicates that magnitude M can be different or
equal to ``0". In Table 1 it can be seen that the relative error for all
of the combinations is bigger than 100\%! In this case the standard
formulas should not be used for a normal distribution, instead more
general statistical formulas should be used.  But this will
considerably increase the width of the confidence interval.  Finally, it
is possible that some deviations of the WEP exist.  Regrettably, the
author has not been able to demonstrate this thesis in his article [1].
The results of the article can only  justify the realization of a new
series of measurements with a more precise  scale and/or a higher number of
the weight measurements for each sample, and a appropriate statistical
processing.

\nonumsection{References}

\end{document}